\documentclass[manuscript,nonacm]{acmart}
\setboolean{@twoside}{false}

\AtBeginDocument{%
  \providecommand\BibTeX{{%
    \normalfont B\kern-0.5em{\scshape i\kern-0.25em b}\kern-0.8em\TeX}}}

\usepackage[
    type={CC},
    modifier={by-nc-sa},
    version={4.0},
]{doclicense}

\begin{document}

\title{Temporally Extending Existing Web Archive Collections for Longitudinal Analysis}

\author{Lesley Frew}
\affiliation{%
  \department{\emph{Department of Computer Science}}
  \institution{\emph{Old Dominion University}}
  \city{Norfolk}
  \state{Virginia}
  \country{USA}
}
\email{lfrew001@odu.edu}
\orcid{https://orcid.org/0000-0003-0929-049X}

\author{Michael L. Nelson}
\affiliation{%
  \department{\emph{Department of Computer Science }}
  \institution{\emph{Old Dominion University}}
  \city{Norfolk}
  \state{Virginia}
  \country{USA}
}
\email{mln@cs.odu.edu}
\orcid{https://orcid.org/0000-0003-3749-8116}

\author{Michele C. Weigle}
\affiliation{%
  \department{\emph{Department of Computer Science }}
  \institution{\emph{Old Dominion University}}
  \city{Norfolk}
  \state{Virginia}
  \country{USA}
}
\email{mweigle@cs.odu.edu}
\orcid{https://orcid.org/0000-0002-2787-7166}


\begin{abstract}

The Environmental Governance and Data Initiative (EDGI) regularly crawled US federal environmental websites between 2016 and 2020 to capture changes between two presidential administrations. However, because it does not include the previous administration ending in 2008, the collection is unsuitable for answering our research question, ``Were the website terms deleted by the Trump administration (2017--2021) added by the Obama administration (2009--2017)?'' Thus, like many researchers using the Wayback Machine’s holdings for historical analysis, we do not have access to a complete collection suiting our needs. To answer our research question, we must extend the EDGI collection back to January, 2008. This includes discovering relevant pages that were not included in the EDGI collection that persisted through 2020, not just going further back in time with the existing pages. We pieced together artifacts collected by various organizations  for their purposes through many means (Save Page Now, Archive-It, and more) in order to curate a dataset sufficient for our intentions. 

In this paper, we contribute a methodology to extend existing web archive collections temporally to enable longitudinal analysis, including a dataset extended with this methodology. We identified the reasons URL candidates could be missing from the initial EDGI dataset, and crawled the past web of 2008 in order to identify these missing pages. We also identified small domains that were vulnerable to being missed by our past web crawl, and found that these domains benefited from a complete web archive index lookup instead. We probed another large collection, the End of Term 2008 dataset, for additional longitudinal candidates, but found that crawler traps were inflating the size of the dataset, leading to only a small number of additional URLs. By analyzing the provenance of the final collection, we determined that this new longitudinal dataset covering three US presidential administrations only exists because of aggregation of artifacts collected by many organizations. We also found that automated brute-force methods alone were not sufficient to create this collection, and that iterative manual analysis of automated results produced more seeds for candidates. Our new dataset includes 1,220 archived triplets (2008, 2016, and 2020) of US federal environmental webpages.

We use our new dataset to analyze our question, ``Were the website terms deleted by the Trump administration added by the Obama administration?'' We find that 81 percent of the pages in the dataset changed between 2008 and 2020, and that 87 percent of the pages with terms deleted by the Trump administration were terms added during the Obama administration. We probed for change trends: when agencies had the same terms repeatedly removed across their websites. We found that certain agencies experienced a large number of change trends, including OSHA, NIH, and NOAA, while 17 of the 30 agencies, including NASA and the Department of Energy, experienced no change trends. Finally, we analyzed the 56 deleted terms and phrases tracked by EDGI and found that the terms fell into two categories: climate and regulation, and identified that there were more change trends in regulation term deletions than climate term deletions.

\end{abstract}






\maketitle

\section{Introduction}

In 2017, the United States Trump presidential administration made international news for deleting large amounts of climate change information from government websites \cite{guardian2017}. 
Trump, a Republican president, took office following Obama, a Democratic president. Is it a valid assumption that the Trump administration was deleting information added to these websites by the previous democratic administration?
It is true that the motivations for removing climate change information are aligned with long-held tenets of the presiding president's Republican political party \cite{grossmann2015ideological}, so examining the webpages only between 2016 and 2020 would not give the full explanations for the deletions. In order to properly examine how the changes on the webpages align with the Republican political party ideals compared to the previous Democratic administration and the next-previous Republican administration, the existing 2016 -- 2020 collection needs to be extended further back in time. 

Nost et al., on behalf of the Environmental Data and Governance Initiative (EDGI), collected and analyzed 40,000 webpages from 2016 -- 2020 in real time \cite{nost2021}. In order to examine this collection further back in time, the prior web archive captures must be added after the fact. Were these webpages captured as far back as 2008, and did they even exist? When certain domains or topics are underrepresented due to this eventuality, how can a collection that spans over ten years be augmented after the events have already occurred? In this paper, we demonstrate techniques that allowed us to extend the EDGI collection to create a longitudinal web archive collection from 2008 -- 2020. This allowed us to analyze the changes on the webpages to determine if the Trump Republican administration was rolling back policies set by an opposing political party between 2008 and 2016, or if the Trump administration was also deleting information from the next-previous Republican administration. This longitudinal analysis allows for contextualization of the webpage edits with regard to the back-and-forth two-party political system in the United States.

\section{Background}

In order to understand the context of this work, we introduce the United States political landscape, which gives clarity to the dates of the mementos in the dataset, along with explanations for the changes on these webpages between different administrations. Next, we introduce the previous work done on the archiving of US federal webpages, to demonstrate the gap in the previous work related to the work we will accomplish. Finally, we present the various provenance sources of archived webpages, including crawling, user initiated captures, and curated collections. All of these types of mementos appear in the final dataset.

\subsection{United States Political Landscape}

The United States political landscape is best characterized as a two-party system with strong polarization. The two main parties are the Republican party and the Democratic party. One of the Republican party's core tenets is limited government, such as through size and less regulation. In contrast, the Democratic party's main ideals include using the government to affect social change and protect environmental regulations \cite{grossmann2015ideological}.

The United States has executive, legislative, and judicial branches, with the president serving as the head of the executive branch. The president is elected every four years, and a president can serve two terms. Table \ref{table:pres} shows US presidents from 1992 -- 2024, corresponding to all presidential administrations represented over the lifetime of the Internet Archive, whose first capture is in 1996. As shown in Table \ref{table:pres}, during this period of time, it is common for the political party of the incoming president to be opposite the outgoing president. The presidential election occurs in November,  and the inauguration occurs the next January. For example, Obama was elected in November, 2008 and was inaugurated in January, 2009. The branches of the US government were codified in the US Constitution; presidential term limits were codified in US Constitution Amendment XXII; inauguration day timing was codified in US Constitution Amendment XX; and election day timing was codified in the  Presidential Election Day Act of 1845.

\begin{table}[ht]
\begin{tabular}{lll}
Election Year            & President Name & Political Party \\
1992 & Clinton & Democratic \\
1996 & Clinton & Democratic \\
2000 & Bush & Republican \\
2004 & Bush & Republican \\
2008 & Obama & Democratic  \\
2012 & Obama & Democratic \\
2016 & Trump & Republican \\
2020 & Biden & Democratic \\
2024 & Trump & Republican            
\end{tabular}
\caption{US Presidents by election year, 1992 -- 2024, showing a frequent switch between the two parties}
\label{table:pres}
\end{table}

%https://en.wikipedia.org/wiki/Wikipedia:Congressional_staffer_edits

The main presidential terms covered in the scope of this paper are Obama 2008 -- 2016, and Trump 2016 -- 2020. Obama, in line with the Democratic party's typical ideals, enacted pro climate change policies \cite{kincaid2013no}. In contrast, Trump's presidency from 2016 -- 2020 was categorized by deregulation \cite{potter2019continuity}, in general and also regarding climate change \cite{jotzo2018us}. This is in line with the Republican party's tenets on small government and lack of environmental regulation.

\subsection{Changes on the Web}

Web pages change over time. Pages are also sometimes entirely deleted from the web. Recent studies show that the median lifespan of a URL is 2.3 years \cite{nypwblog}. Users access webpages via a browser, which relies on Hypertext Transfer Protocol (HTTP). 
HTTP transmits the page content as well as a status code to indicate success \cite{fielding2022rfc}. 
Successful requests have a 200 OK HTTP status. Pages that redirect have a 3xx status code. When the HTTP request cannot be fulfilled, a 4xx status code for a client-side error is returned. Pages that are not found, for instance, if they have been deleted, typically have a 404 status code. Due to the ephemeral nature of the Web, web archives preserve historical copies of webpages and provide access to users to view these past versions.

Multiple similar URLs can resolve to the same webpage, such as with or without www. Since the URL is the key used for lookup in a web archive, a canonicalization technique called Sort-friendly URI Reordering Transform \cite{surt}, or SURT, is used to match these multiple versions of a URL. While users may perceive webpages with different URLs beyond canonicalization to be the \textit{same} webpage,\cite{klein2014moved} such as in the case of a machine readable redirect, or in the case that the page is moved but no machine readable redirect is left behind, we consider these webpages to be out of scope for the purpose of this study. Current web archive lookup and change analysis tools rely on webpages having the same canonical URL over time, so we restrict our dataset to just those pages whose URL persisted from 2008 through 2020 without changes beyond the canonicalization rules.

\subsubsection{Archiving of US Federal Websites}

Nost et al. \cite{nost2021}, on behalf of the Environmental Data and Governance Initiative (EDGI), monitored changes on 30 US federal environmental agency websites between 2016 and 2020. They compared the change in 56 pre-chosen environmental terms and phrases on 40,000 webpages using the web archive holdings at the Internet Archive. 

The archival of federal websites, especially at the end of a president's term, is an important task undertaken by multiple organizations. The End of Term Web Archive is created through a partnership between five organizations, including the Internet Archive and the Library of Congress 
\cite{seneca2012takes, phillips2017end}.
This web archive includes a full-text search feature,\footnote{\url{http://eot.us.archive.org/search/}}
but each end of term crawl includes only one capture of each web page. Phillips et al. \cite{phillips2016exploratory} compared the 2008 and 2012 end of term collections to identify changes in crawl dates and webpage addresses, but individual terms were not analyzed. Our previous work with the EDGI dataset studied how to more easily find the deleted terms within web archives \cite{frew2023} and how the terms deleted related to actual user search query terms for those pages \cite{frew2024}.

\subsubsection{Editing Motivations}

There are neutral reasons for human editors to make changes to webpages. One reason is if the information on the page becomes stale. The human editor may notice, even perhaps from another webpage, that the information on their webpage is no longer up to date. They will make changes so that the website content is brought back up to date \cite{adar2009temporal}.

On Wikipedia, Yang et al. \cite{yang2017identifying} identified common intentions that motivated edits. Some are neutral, such as updating information as stated above, copy editing and refactoring, and adding citations. Other types of edits actually serve to make the article more neutral, such as editing the point of view or removing information that is untrue. Editors will also add information that supports the content already on the page, clarify existing information, or simplify existing verbiage.

There are also multiple reasons that are not neutral that cause human editors to make changes on webpages. One reason is to improve the ranking of the webpage on search engine results pages. This is referred to as rank incentivized document manipulation, or search engine optimization \cite{kurland10.1145/3477495.3532771}. Another reason that human editors make non-neutral changes to webpages is to make the webpage less neutral, which is called tendentious editing. Groups of people who are known to benefit from furthering a specific view point on a controversial subject include politicians promoting their party's viewpoints, businesses promoting their products, and others.

\subsection{Provenance of Archived Webpage Collections}

\begin{figure}[ht]
  \centering
  \includegraphics[width=\linewidth]{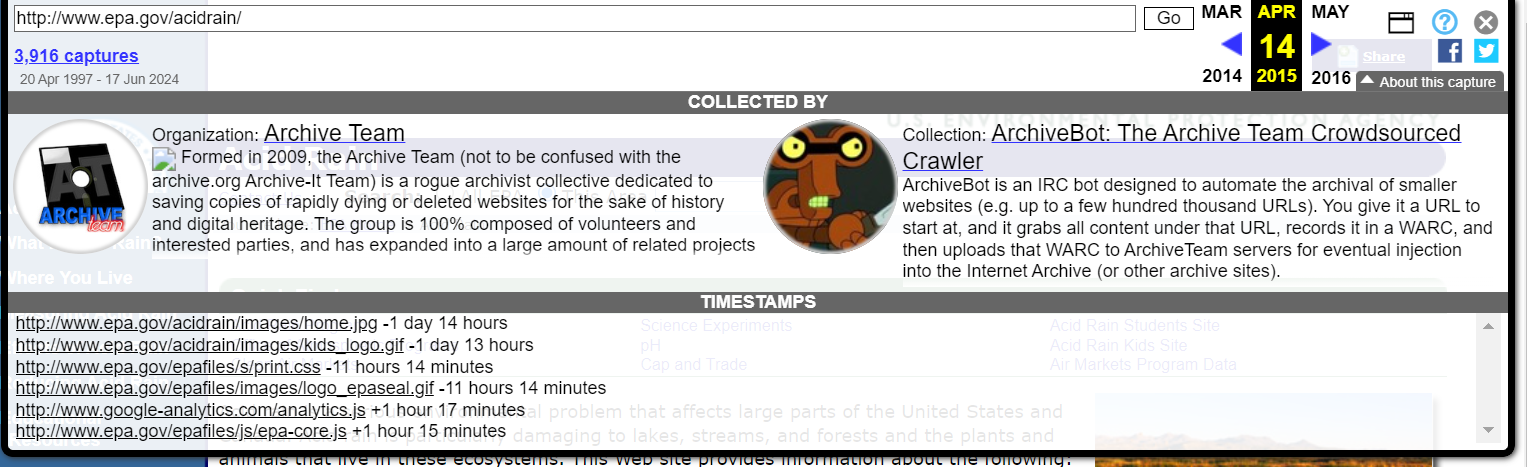}
  \caption[Wayback Machine Provenance Information]{Wayback Machine Provenance Information for epa.gov/acidrain from 2015-04-14 showing the capture source as Archive Team}
  \label{fig:prov}
\end{figure}

When using web archives as data, the source of the data is important to consider. For example, Taylor \cite{taylor2023} identified the source, or provenance, of an archived web page to be extremely relevant information when the archived web page is used in legal proceedings. Researchers are also interested the provenance of pages in a collection. Ben-David et al. \cite{ben2018internet} found that the collection of North Korean websites they analyzed consisted of crawls from various web archives, commercial organizations like Alexa, national organizations and libraries, grassroot organizations like Archive-Team, and personal web archive collections crawled using Archive-It. Figure \ref{fig:prov} shows a webpage captured by Archive team, as displayed in its provenance section at the Internet Archive's Wayback Machine. It is this combination of crawls, citizen archiving, and aggregation from many organizations that result in curated collections after the fact.

%https://web.archive.org/web/20231129001829/epa.gov/acidrain

\subsubsection{Crawling}

In the same way that search engine bots crawl the web to keep their index fresh, web archives rely heavily on data gathered through crawlers. 

Existing data sets of webpage crawls commonly used for academic purposes, such as ClueWeb \cite{overwijk2022clueweb22} and Common Crawl,\footnote{\href{https://commoncrawl.org/}{https://commoncrawl.org/}} aim to collect snapshots of a large amount of unique URLs. Other large crawls focused on specific domains include the End of Term Archive \cite{seneca2012takes, phillips2017end, phillips2023}. Many of the strategies used for live web crawling, such as by search engines, are also necessary when crawling for web archives. One example is avoiding crawler traps \cite{bickford2023sos,maemura2023sorting}, which are potentially unbounded subpages in websites such as calendar pages. 

The Internet Archive conducts many different types of crawls. \cite{ben2018internet}. A \emph{wide crawl} aims to collect as many pages, both \emph{high} and \emph{deep}, as possible. A \emph{survey crawl} aims to collect one top page from as many domains as possible. The Internet Archive uses many seeds, including Wikipedia outlinks as well as manually curated seeds for the other crawls. These kind of crawls are used for a comprehensive collecting strategy \cite{pastweb}.

When a more focused past web collection is required, different crawling strategies are needed. 
Most event collections are crawled at the time of the event. Web crawlers like Heritrix \cite{mohr2004introduction} are available for crawling the live web for the purpose of archiving the found pages. However, less work has been done on crawling the past web. Klein et al. \cite{klein2018focused} crawled the past web without restrictions on the target domains to build past web event collections. In order to restrict the web archive to a set time, they used a sticky time policy rather than a sliding time policy employed by past web replay systems \cite{ainsworth2013evaluating}. Gossen et al. \cite{gossen2020towards} crawled only outlinks that were predicted to be relevant to the target collection, taking into account both content and time \cite{pastweb}. 

Ben-David et al. \cite{ben2018internet} explain the benefits and downsides of relying on crawling. One of the benefits is that crawlers may find important pages by accident in the course of their exploration. These pages were not known to be important at the time they were archived, but meaningful changes on the pages later or a full page deletion caused the earlier capture to be important, too. Another benefit of crawling is scale. Crawling results in more data than humans could generate manually. The downsides of relying on crawling is that there is no guarantee any single page will be put onto the crawl horizon, which means there is no guarantee of any single page being archived consistently via multiple crawls. Pages that do not have many links to them are unlikely to be discovered via crawl. Crawling requires humans to determine effective seeds for the initial page visits \cite{ben2018internet,pastweb}.

%Frew et al. \cite{frew2023} found that US federal environmental websites archived in 2016 consisted of crawls from national (Library of Congress) and international (Portuguese Web Archive) web archives. 

%Frew et al. CITE CIKM ARXIV found that US federal environmental websites archived in 2008 consisted of crawls from commercial organizations like Alexa, non-profit organizations like Common Crawl, national and international (INA French archive) web archives, research organizations like the End of Term Archive, public and private universities, state government agencies, state libraries, and K-12 organizations.

\subsubsection{Save Page Now}

\begin{figure}[ht]
  \centering
  \includegraphics[width=0.9\linewidth]{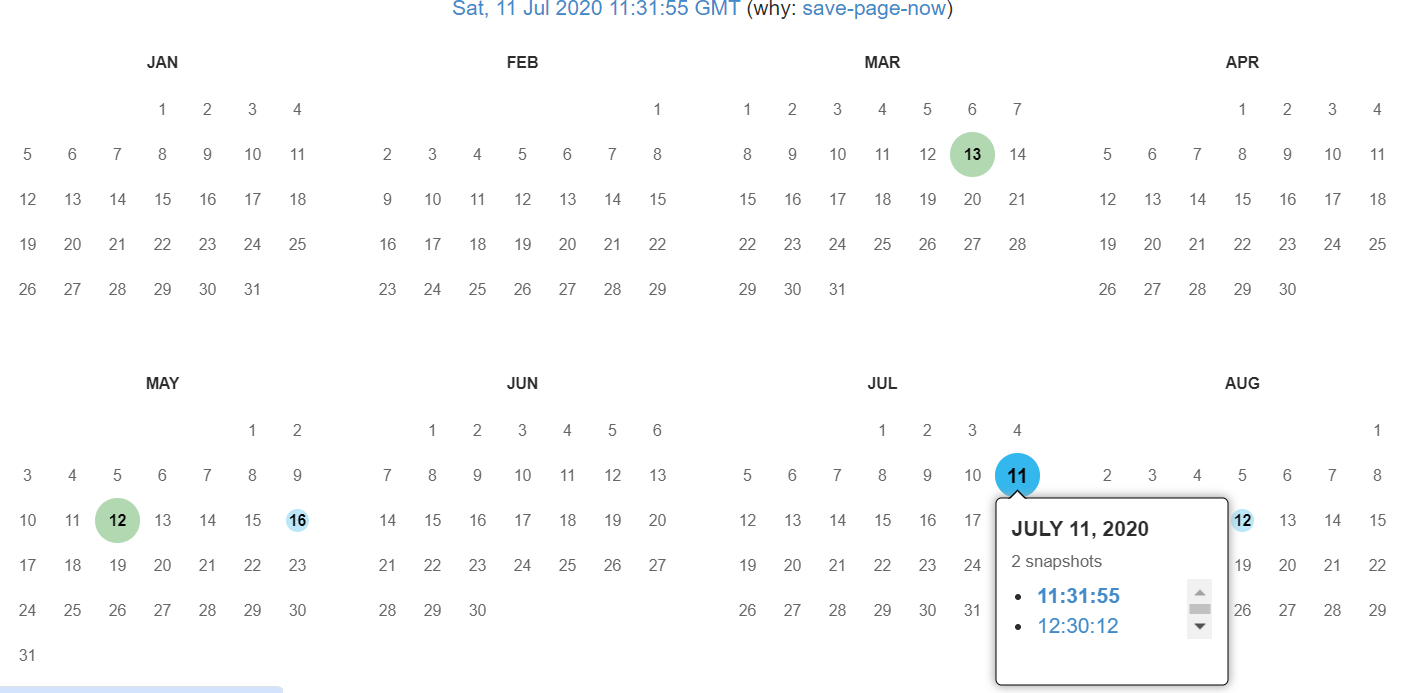}
  \caption[Save Page Now for georgeforny.com]{Someone used Save Page Now to archive a copy of georgeforny.com on 2020-07-11, a page known for contentious and frequently changing information.}
  \label{fig:prov3}
\end{figure}

Save Page Now is a feature at the Internet Archive that allows users to request an immediate archival of a webpage. The feature has been available since 2014 \cite{ben2018internet}. While crawling requires collecting many resources and analyzing the collection after the fact for meaningful captures, pages archived with Save Page Now were already deemed meaningful at the time of capture. One reason why users deem captures meaningful is because they think the information will be changed or the page will be deleted \cite{hanson2022preserving,rakityanskaya2023belarusian}. In Figure \ref{fig:prov3}, someone used Save Page Now on georgeforny.com, which is a page with information that is contentious and frequently changes \cite{frew2023santos}.

%https://web.archive.org/web/20200801000000*/georgeforny.com

\begin{figure}[ht]
  \centering
  \includegraphics[width=0.9\linewidth]{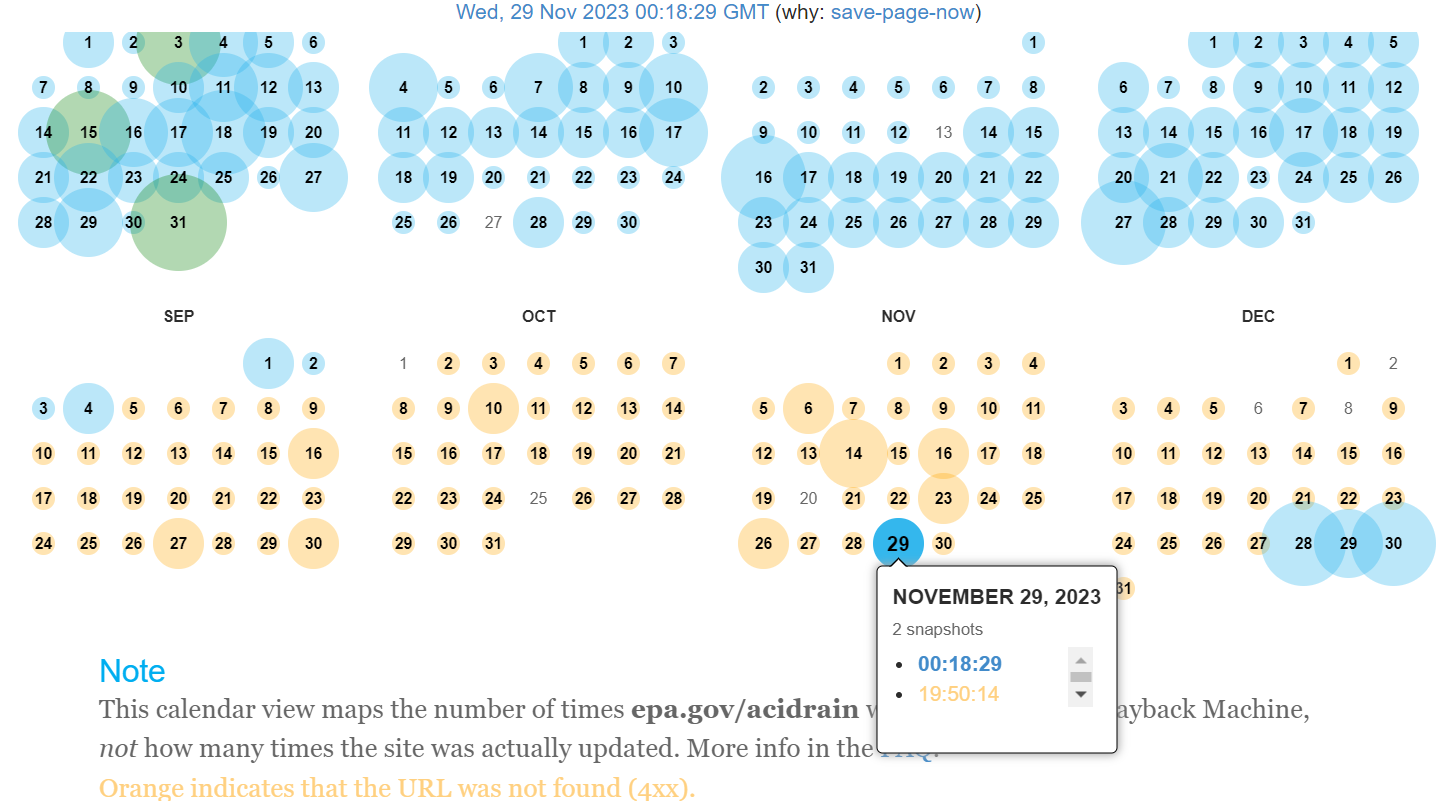}
  \caption[Save Page Now for epa.gov/acidrain]{Someone used Save Page Now to successfully archive a copy of epa.gov on 2023-11-29. The mementos being saved via crawling in the same time period were not successfully captured.}
  \label{fig:prov2}
\end{figure}

Ogden et al. \cite{ogden2024know} found that human initiated Save Page Now requests resulted in higher fidelity captures than bot initiated requests. In Figure \ref{fig:prov2}, someone used Save Page Now on epa.gov/acidrain because the other mementos being collected via crawl were not archiving successfully.

%https://web.archive.org/web/20230701000000*/epa.gov/acidrain

\subsubsection{Curating Collections}

Web archives can be affiliated with a country (Portuguese Web Archive, Library of Congress, and so on) or established independently (Internet Archive, perma.cc),\cite{Gomes:2011:SWA:2042536.2042590} and these web archives are affiliated through the International Internet Preservation Consortium.\footnote{\url{https://netpreserve.org/about-us/members/}} Additional organizations have a need for the use of these web archives\cite{2022-ndsa-web-archiving-survey}. Organizations whose scope is beyond web archiving that wish to perform their own crawls and curate their own smaller web archive collections can use Archive-It to accomplish this goal. Rakityanskaya \cite{rakityanskaya2023belarusian} detailed how academic organizations along with community grassroots organizations used Archive-It to document Belarusian political activism on a variety of websites including media outlets as well as personal blogs. Archive-It collections can also be combined to form larger collections. Mark Graham, the director of the Wayback Machine, has curated intriguing collections that have been used to form larger collections on topics such as North Korea \cite{ben2018internet}, Belarus \cite{rakityanskaya2023belarusian}, and CNN \cite{weigle2023right}.
%Frew et al. (cite cikm) found that these organizations can include public and private universities, state government agencies, state libraries, and K-12 organizations.

Beyond establishing their own seeds for crawls, users also wish to find the most salient captures within the crawls they or someone else has performed. The entire crawl is typically filled with pages that are not meaningful, or no clear entry point is clear. Choosing the best representative pages to summarize a collection is one way to address this problem \cite{jones2021improving}. Some collection curators also prefer manual curation over focused crawls \cite{pastweb,rakityanskaya2023belarusian}.

\section{Method}

In order to temporally extend an existing web archive collection, the existing links must first be analyzed to identify initial candidates with large temporal spans. In order to find more candidates beyond the initial list, the collection should be analyzed for potentially missing pages. Crawling the past web can surface additional candidates, but the results of the crawl should also be analyzed to identify candidates missed by the crawl. Additional collections with similar scope can be combined with the original collection to find more candidates. Finally, manual augmentation based on analysis can sometimes produce candidates faster than automated brute-force means. Using this method, our goal was to create a dataset of US federal environmental webpages that persisted from 2008 to 2020, with 15 high and 15 deep links from each of 30 agencies identified by EDGI. After producing the dataset, it should be analyzed for changes to determine if the administration in 2020 was deleting terms added by the previous administration.

\subsection{Extracting Candidates from the Original Collection}

The first step in extending the EDGI environmental changes dataset back to 2008 is to analyze each existing page in the current dataset that resolved successfully in both 2016 and 2020 to see if it also resolved successfully in 2008. This involves two steps: the first step is to see if there are any archived versions of these pages at any web archive in 2008, and the second step is to check the archived HTTP status of each page with a 2008 memento.

In order to query multiple web archives about their holdings for these specific pages, we used Memgator \cite{alam2016}. MemGator aggregates the results of the query for each page into a TimeMap\cite{van2013http}. The EDGI dataset was separated into two parts: approximately 10,000 pages contain paired mementos from 2016 and 2020 at the Internet Archive, and the other 30,000 pages did not fit this criteria. We collected those 30,000 TimeMaps using MemGator in fall 2022 \cite{frew2023}, and found additional paired mementos at other web archives besides the Internet Archive. Particularly, the Library of Congress web archive contained additional mementos not present at the Internet Archive for these government pages. For the other 10,000 mementos, we collected those TimeMaps in fall 2023. While previously the Library of Congress web archive was a significant source of additional mementos for 2016, it did not contain any significant number of added mementos beyond the Internet Archive's holdings for 2008.
Table \ref{table:edgi30ksource} shows the count of mementos for the 30,000 TimeMaps collected in fall of 2022 when the Library of Congress web archive and UK Web Archive were still accessible to the public. We also analyzed which, if any, of the 30,000 pages had no mementos at the Wayback Machine in 2008. We found only four such pages with mementos in an archive other than and excluding the Wayback Machine, each only at Archive-It; however, four out of thirty-thousand is not a significant percentage.
Therefore, we restricted our analysis to the Internet Archive.

%https://github.com/lesleyodu/odu_phd/blob/main/2025-04-15/timemapout.txt
\begin{table}[]
\begin{tabular}{lr}
Web archive            & 2008 memento count for EDGI 30k \\
web.archive.org        & 8919                       \\
wayback.archive-it.org & 1170                       \\
webarchive.loc.gov     & 269                        \\
arquivo.pt             & 7                          \\
web.archive.org.au     & 2                          \\
swap.stanford.edu      & 1                          \\
archive.md             & 1                          \\
wayback.vefsafn.is     & 1                         
\end{tabular}
\caption{Count of mementos from 2008 at each web archive represented in the 30,000 TimeMaps collected in fall of 2022 for the EDGI dataset.}
\label{table:edgi30ksource}
\end{table}

The Internet Archive offers a CDX API\footnote{\href{https://github.com/internetarchive/wayback/tree/master/wayback-cdx-server}{https://github.com/internetarchive/wayback/tree/master/wayback-cdx-server}} which contains a list of all mementos for a given URI and the archived HTTP status codes for those mementos. We downloaded the CDX file for each candidate page with successful mementos in 2016 and 2020, and a candidate memento in 2008, to get the status of the 2008 memento. Almost all of the candidates had a 2008 memento with a successful (200) archived HTTP status in 2008. While having a successful memento in both 2016 and 2020 did not reliably predict whether or not a page would have a memento in 2008, having successful mementos in both 2016 and 2020 and having at least one memento in 2008 was a good predictor of the 2008 memento being successful also. We started with 10,673 candidates that had successful paired mementos in 2016 and 2020, and 961 had a candidate 2008 memento, so 9\% became candidates for an archived triple. 62 were further eliminated, but this percentage (6\%) is small. Therefore, the main reason a pair would not extend back to 2008 was a lack of archived capture for 2008, either because the page did not exist or because it was not crawled. In all, there are 899 2008--2016--2020 triplets originating from the EDGI 2016-2020 pairs dataset.

%idk where these numbers below are from
%10673 = from jcdl sankey
%961 from Progress Report 2023-11-21 to 2023-11-27
%899 from final triplet dataset fromedgi = true
%40,000 URLs to 11,000 2016/2020 paired mementos to 1122 with at least 1 memento before 2008. 1076 out of 1122 candidates had a successful 2008 memento. Then filtered on 30 agencies to get abc.

\subsection{Identifying Attributes of Extension Candidates}

The EDGI environmental changes dataset focused on 30 government agencies. \cite{nost2021} It also differentiated between high level pages and deep pages, and used this property in its analysis of changes on the pages. Table \ref{table:edgi2008only} shows the distribution of the 899 triplets by agency and path depth using only EDGI as a source. 

\begin{table}[]
\begin{tabular}{lll}
Domain              & 2008 EDGI high & 2008 EDGI deep \\
blm.gov             & 1              & 0              \\
cdc.gov             & 48             & 5              \\
doi.gov             & 7              & 0              \\
dot.gov             & 2              & 0              \\
energy.gov          & 2              & 0              \\
epa.gov             & 66             & 3              \\
federalregister.gov & 1              & 0              \\
\textbf{ferc.gov}   & \textbf{114}   & \textbf{67}    \\
fws.gov             & 37             & 12             \\
gao.gov             & 1              & 0              \\
globalchange.gov    & 2              & 0              \\
hhs.gov             & 2              & 0              \\
justice.gov         & 1              & 0              \\
nasa.gov            & 4              & 0              \\
\textbf{nih.gov}    & \textbf{26}    & \textbf{50}    \\
\textbf{noaa.gov}   & \textbf{198}   & \textbf{56}    \\
nps.gov             & 3              & 0              \\
\textbf{osha.gov}   & \textbf{98}    & \textbf{62}    \\
osmre.gov           & 1              & 0              \\
usda.gov            & 4              & 1              \\
usgs.gov            & 5              & 0              \\
whitehouse.gov      & 3              & 0             
\end{tabular}
\caption{Count of triplets by agency and path depth from EDGI}
\label{table:edgi2008only}
\end{table}

While the 40,000 pages in the original dataset were important enough to track explicitly by EDGI, there are many pages they did not track but were present on the 30 agencies' sites, demonstrated by the fact that there are 15.8 million URLs for these domains in the 2008 end of term crawl. The pages explicitly tracked by EDGI were seeded by humans observing change in real time on those sites. EDGI acknowledged the limit of their crawler as well at 50 levels of depth. EDGI did not track the additional pages on the agency websites because they were not observed to be having changes, they could not be crawled from the seeds at all, or the crawler timed out before getting to the additional pages. Unless additional pages are analyzed, it will not be possible to tell the difference between these three possibilities. Therefore, it is within scope to analyze additional mementos that may have experienced change but might have been missed by the EDGI crawler. 

Optimally, the dataset will contain 15 high and 15 deep triplets per agency. However, Table \ref{table:edgi2008only} shows that only using EDGI URLs results in only four agencies meeting this criteria. In addition, domains included in EDGI such as fema.gov, gsa.gov, and nsf.gov had no links in 2008 through this method, even though those domains did exist in 2008.

To find these additional mementos, some strategies include past web crawling, full domain analyses for small domains, data collection augmentation, and manual augmentation. These strategies are detailed further below.

\subsection{Crawling in 2008}

%https://github.com/lesleyodu/odu-ms-thesis/blob/main/pr/2023-03-21/edgi_crawl.py

To find candidates using the EDGI dataset, we started with candidates that had successful HTTP status codes in 2020 and 2016 and checked to see if they extended back to 2008. Our next strategy used the opposite approach: find candidates that had successful HTTP status codes in 2008 and check to see if they extend forward to 2016 and 2020.

Most event collections are crawled at the time of the event. Web crawlers like Heritrix CITE are available for crawling the live web for the purpose of archiving the found pages. However, less work has been done on crawling the past web. Klein et al. \cite{klein2018focused} crawled the past web without restrictions on the target domains to build past web event collections. In order to restrict the web archive to a set time, they used a sticky time policy rather than a sliding time policy employed by past web replay systems \cite{ainsworth2013evaluating}. Because our collection has already been built once in the past, and the resulting collection had restrictions on the possible allowed domains, we manually crawled the past web using the sticky time policy and only kept links on each page for .gov domains. This eliminates the possibility of finding government web pages that were linked to externally, but keeps other government websites beyond the 30 agencies as possibilities for internal links. The sticky policy meant that each 2008 link was requested for January 1, 2008, and if it did not have a successful archived HTTP status or if the datetime of the memento returned was not in 2008 or 2007, then the crawler returned from that page to the rest of the horizon. 

%https://github.com/lesleyodu/odu-ms-thesis/blob/main/pr/2023-09-18/stats.csv

We ran the crawler two times with different seeds based on which agency websites were not represented in the first crawl. Even after repeated calls, it became clear that some domains were either not well archived or not well connected because they had none or a small number of mementos returned by the crawls. We determined that four of the domains (bia.gov, boem.gov, eia.gov, and fs.usda.gov) for the agencies did not exist in 2008, which explains why there were no mementos for those pages in 2008. We were able to estimate the prevalence of mementos for the other domains by examining the End of Term Web Archive 2008 counts for the domains. \cite{phillips2023}. 

%https://github.com/lesleyodu/odu-ms-thesis/blob/main/pr/2023-11-27/get_more_candidates_crawl.py
After obtaining the list of 2008 candidates for each domain, we used the same strategy as we used on the EDGI elements to analyze the CDX file for possible mementos in 2016 and 2020. We had 79,791 candidates identified from the crawl. However, we did not check every candidate; instead, we examined candidates by agency and checked high path depth links first and stopped once we found 15 or ran out of high links, and the same for deep path links. For example, usgs.gov had 2,743 crawl candidates but once we found 15 of each the script terminated. We conducted this process separate from the EDGI candidates, so some agencies ended up with slightly above or below 15 candidates. For example. cdc.gov had 5 deep candidates from EDGI as shown in Table \ref{table:edgi2008only}. Examining crawl candidates led to 15 deep links, but because of overlap with the EDGI candidates, the final dataset contains 18 deep links for cdc.gov.

While these links were generated because they existed in 2008, there was no guarantee they would persist through 2020. This was the most common reason a 2008 candidate would not qualify as a triplet. For example, we had 45,797 crawl candidates from whitehouse.gov, 1,433 of which were high path depth and the rest deep. We ran all high candidates, though all 3 found were already found through EDGI, and after 24 hours of running deep candidates we had yielded none with a successful 2020 status code. whitehouse.gov is clearly more susceptible to deleting pages between administrations than even the other agencies, because of its direct tie to the current president. 

The next most common reason was that a candidate had a memento with a successful HTTP status code in 2017 or 2021 for example, but was missing 2016 or 2020. For some of these candidates, there is evidence from other pages in the same directory that the pages were fully deleted between 2016 and 2020 but were later restored. One example of this is http://www.nps.gov/history/archeology/EAM/landmarks.htm, which was archived more frequently than the other pages in that subfolder. For other candidates, all of the other mementos for that page and for pages in the same directory have successful HTTP archived status codes. Therefore, those pages were not being archived with enough frequency to be analyzed as a triplet. One example of this is http://www.nps.gov/history/archeology/aiassess/index.htm which has a successful memento in 2017 and 2022 but not in 2020. The Wayback Machine has increased the count of mementos crawled by a power of ten for each administration, so in 2008 there were just exponentially fewer mementos and that many fewer chances that pages of interest were captured.

%power of 10 comes from edgi 30k timemaps

%some examples
%https://docs.google.com/document/d/1bHew7DrBw5PZSVlgHBJz-aEr3clXYLDJBuoA8bUr8fo/edit?tab=t.0

\subsection{Full domain analyses}

While crawling the past web, 3 domains had few to no resulting candidates from the crawl, but web archive queries indicate the domains were in existence in 2008. Additionally, the three domains have non-zero mementos archived in the 2008 End of Term archive. These domains were federalregister.gov, globalchange.gov, and osmre.gov. For these domains, we did a prefix CDX query restricted to 2008. These CDX queries ranged from 38 pages to 600 pages. Due to rate limiting at the Internet Archive aimed to combat crawling for LLM training purposes, we had to modify our original methodology \cite{weigle2023right} to add a delay of between 8 and 11 seconds between requesting CDX pages. We also then crawled all pages found with a depth of 1 to discover additional pages linked and listed in navgigation sections.
%update 2025 rate limiting 8-11 seconds can cite new blog
%Due to rate limiting at the Internet Archive aimed to combat crawling for LLM training purposes, we could only request one CDX page per 5 minutes. Therefore, this strategy is well suited for domains with a small number of mementos in the time range, but is prohibitive for larger domains. 

For federalregister, while the single URL representing the root domain was able to be added to the triplets list, no other URLs in this domain had mementos until 2011. For globalchange, we added 3 new URLs to the triplets list. Additional URLs, found via crawling, existed before 2008 but had no 2008 mementos. Other URLs in the same folders persisted through 2008, so it is more likely that these URLs also persisted through at least 2008 but were not crawled rather than that they had a non-successful memento in 2008. For example, http://www.globalchange.gov/policies/comments/0011.html has no memento in 2008, but 0001 through 0012.html are in the 2008 CDX with successful 200 mementos. However, the files in this subfolder do not have mementos beyond 2008 so they were not added to the triplets list. For osmre.gov, we added 3 new URLs to the triplets list. The subsequent crawl resulted in one additional candidate subdomain beyond the CDX query. We used a prefix CDX query, which does not return subdomains. A different query, a domain query, does return subdomains and that would be the preferred CDX query method for future researchers due to this example. However, the subdomain techtransfer.osmre.gov did not persist beyond 2014. 

Another example is epa.gov. The 2008 past web crawl resulted in 895 candidates for deep links for this domain, and six deep links in the `ttn' subfolder turned up as positive hits for triplets. However, the crawl was just a sample compared to the entire holdings, and we suspected there might be more pages in the `ttn' folder that were positive triplets. We then looked for `ttn' subfolder hits in the 2020 CDX, since we knew this folder persisted back to 2008 but only some pages would have persisted forward, and there were just a few hundred to check. Using the CDX from 2008, 2016, and 2020 for these pages, we identified many more triplets.

\subsection{Combining Collections from Different Time Periods}

While the EDGI environmental changes dataset was collected between 2016 and 2020, there are additional datasets that include government webpages collected during earlier time periods. The End of Term Web Archive is created through a partnership between five organizations, including the Internet Archive and the Library of Congress \cite{seneca2012takes, phillips2017end}. The End of Term collected data from 2004 through 2020 has been made available \cite{phillips2023}. Thus, this dataset collection does include the target time period of 2008. In fact, there are 15.8 million URLs from the 30 EDGI agencies in the end of term 2008 crawl.

Unfortunately, many of these candidates did not persist with a successful HTTP status code through 2020. In addition, the number of URLs in the end of term dataset is inflated due to crawler traps. For example, approximately 75\% of the end of term crawl for the Bureau of Land Management is crawler traps. This domain was of particular interest for use with the end of term crawl because neither EDGI nor the 2008 past web crawl produced more than a few triplets, but end of term had 280,000 candidates for this domain. Another issue is that the large number of candidates also makes it prohibitive to check all of them, compared to EDGI and the 2008 past web crawl which had fewer than 100,000 candidates for all domains combined. blm.gov appeared to have been reorganized in 2016, preventing the existence of most 2008--2016--2020 persistent triplets. The globalchange.gov domain also had similar issues to blm.gov, in that there were crawler traps also few links from the other methods. For nih.gov, the only additional links we found from the 2008 end of term dataset were subdomains.

Due to the same issue with the 2008 past web crawl, that many of these links did not persist forward in time to 2020, along with the crawler traps, the 2008 end of term dataset did not contribute a significant amount of new triplets to the final dataset. Further analysis of the provenance shows that 65 of the 1,220 mementos from 2008 were archived due to the end of term crawl. This suggests that the end of term crawl is a better source for historians who wish to analyze pages that have been removed from federal websites rather than pages that have persisted.

%add this table here before manual
%energy.gov find eot count or redo - found
%https://github.com/lesleyodu/odu-ms-thesis/blob/main/pr/2023-09-18/stats.csv

\subsection{Manual Augmentation}

\begin{figure}[h]
  \centering
  \includegraphics[width=0.9\linewidth]{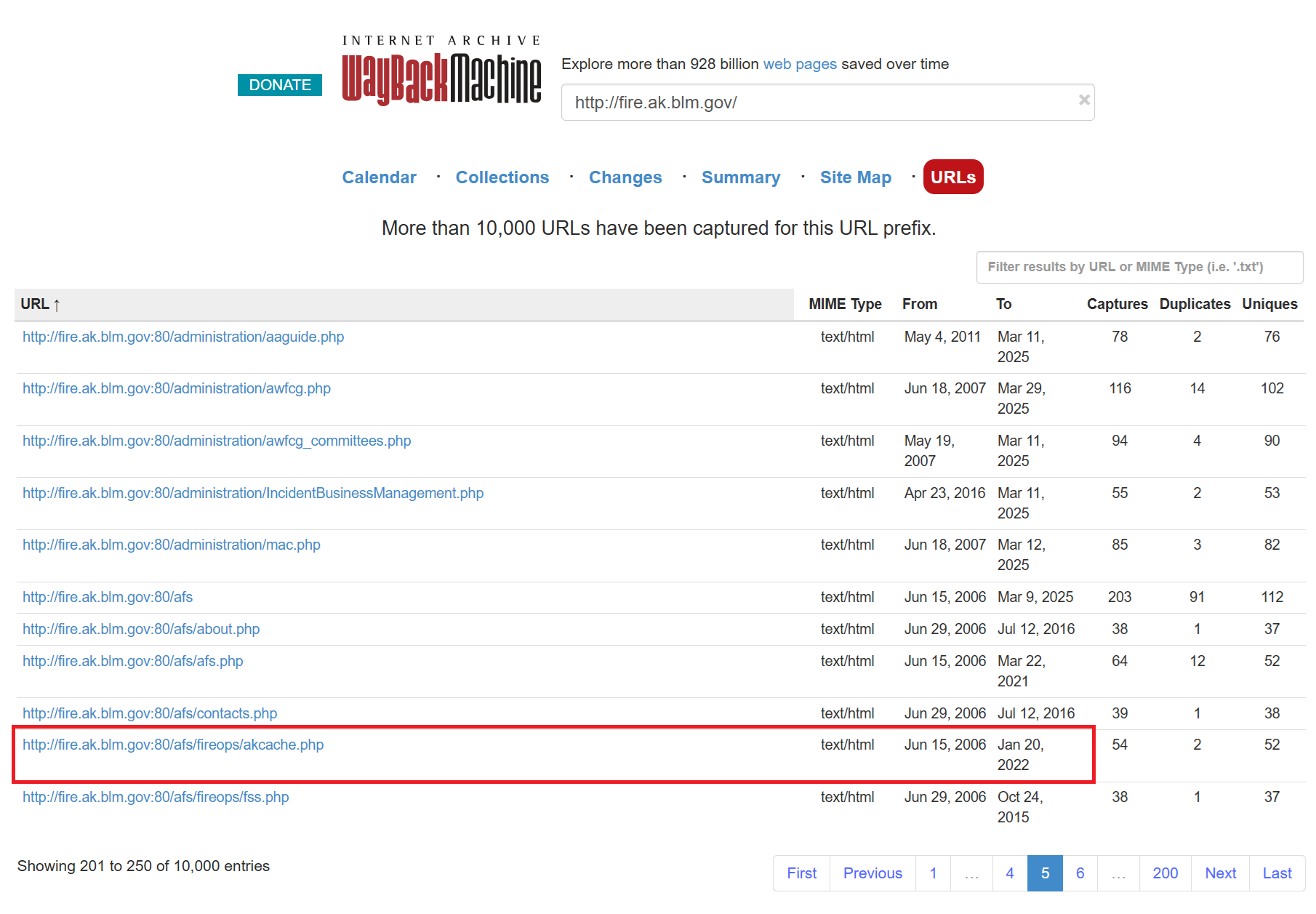}
  \caption{Wayback Machine URLs tool: the first capture of the akcache page was before 2008 and the last capture was after 2020, indicating it should be examined further}
  \label{fig:manualcandidate}
\end{figure}

Even with all of the above strategies, manual augmentation proved to be one of the most successful ways to find additional candidate URLs beyond EDGI. For domains that had few results with the other methods, the following manual methods were employed. First, we did a prefix query using the web interface for each successful triplet's directory using the URLs tool. If a directory was deep, we also queried upper level directories. The web interface gives the earliest and most recent archival dates of each page, , as shown in Figure \ref{fig:manualcandidate}, which made it easy to find candidates archived in 2008 and 2020. Then, manual inspection of the web interface TimeMaps allowed for quick elimination of candidates with non-successful HTTP statuses, since these are colored differently on the web interface, as shown earlier in Figure \ref{fig:prov3}.

%\begin{figure}[h]
%  \centering
  %\includegraphics[width=0.9\linewidth]{Figures/manual3.png}
  %\caption{Wayback Machine calendar UI captures listing page with colors indicating HTTP status: this page's capture in 2020 is in yellow, indicating it was not successful.}
  %\label{fig:manualcoffeestain}
%\end{figure}

\clearpage

\begin{figure}[h]
  \centering
  \includegraphics[width=0.8\linewidth]{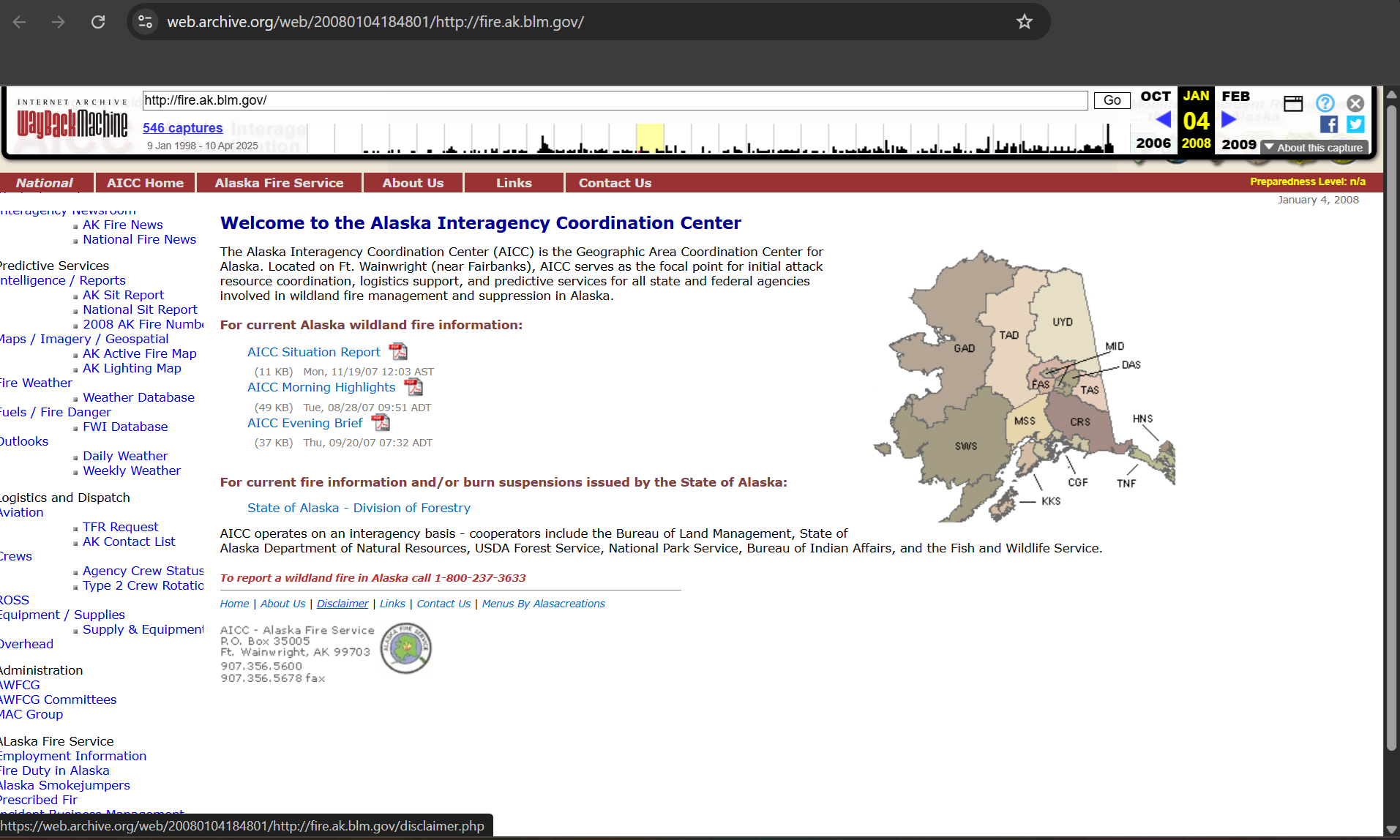}
  \caption{Replaying the page fire.ak.blm.gov in 2008 via the Wayback Machine yields links to other triplets.}
  \label{fig:manualreplay}
\end{figure}

The other manual strategy employed was to replay any positive results from each domain in the past web and manually click on links on the page to see if they were also available in 2008, and then into 2020. This is similar to doing a past web crawl of depth 1. This strategy is shown for the page fire.ak.blm.gov in Figure \ref{fig:manualreplay}.

\begin{figure}[h]
  \centering
  \includegraphics[width=0.8\linewidth]{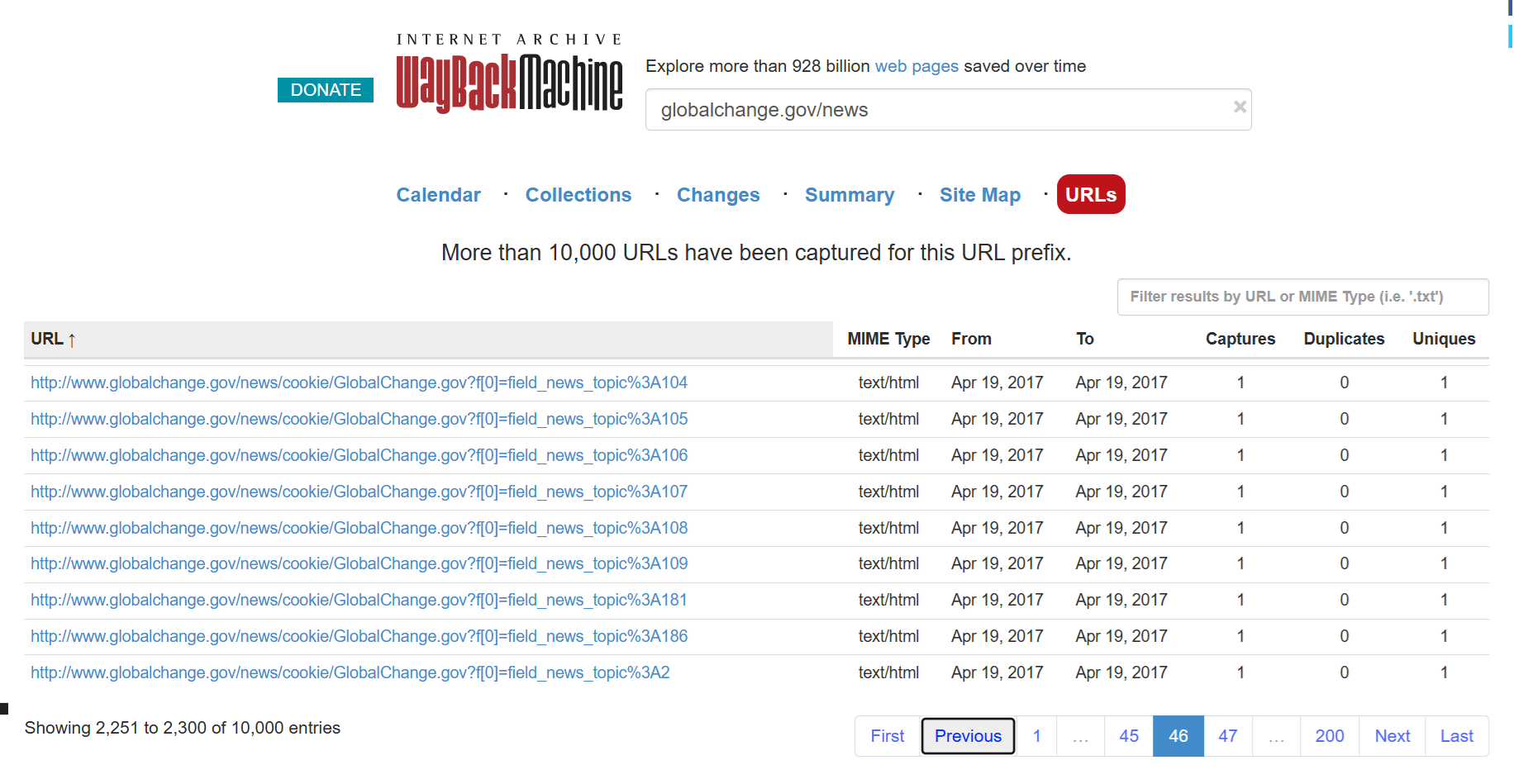}
  \caption{Wayback Machine URLs tool with crawler trap: pages 4 through 46 all contain captures in this `cookie' directory}
  \label{fig:manualcrawltrap}
\end{figure}

One difficulty with this method is the crawler traps, as shown in Figure \ref{fig:manualcrawltrap}. This figure shows that 25\% of what is shown to the user for the globalchange.gov/news directory URLs tool would not yield pages for analyzing.

%individual domain case studies - incorporated
%https://docs.google.com/document/d/1DMuWUbysoxMEGAZD_p4qWGkkw5d4aU9JroFlx_nNfE8/edit?tab=t.0

\clearpage

\section{Results}

In this section, we present our final dataset of 1,200 2008--2016--2020 archived triplets. We also examine the provenance of the new 2008 pages and how that differs from the later years. We also address our research question to determine if the terms deleted in the Trump administration were added in the previous democratic administration. We find a stronger pattern of regulatory term deletions than climate change deletions, and demonstrate that a vast majority of the pages with deleted terms had those deleted terms initially added during the Obama administration.

\subsection{Extended Collection}

%https://observablehq.com/d/d31b149154c9fa80

% high deep by tld

Table \ref{table:final} shows the final distribution of triplets in the dataset by agency and path depth. Seven agencies met the criteria for 15 links at both high and deep path depths. Another five agencies had at least 15 links in one category. 15 agencies had fewer than 15 links in total, and 15 agencies had no deep links. 

\begin{table}[ht]
\begin{tabular}{lll}
  domain                  & high & deep \\
blm.gov             & 5    & 2    \\
cdc.gov             & 50   & 18   \\
doe.gov             & 3    & 0    \\
doi.gov             & 10   & 0    \\
dot.gov             & 3    & 0    \\
energy.gov          & 2    & 0    \\
epa.gov             & 108  & 52   \\
fed.us              & 3    & 0    \\
federalregister.gov & 1    & 0    \\
fema.gov            & 1    & 0    \\
ferc.gov            & 114  & 67   \\
fws.gov             & 37   & 53   \\
gao.gov             & 27   & 0    \\
globalchange.gov    & 3    & 0    \\
gsa.gov             & 2    & 0    \\
hhs.gov             & 12   & 0    \\
justice.gov         & 20   & 4    \\
nasa.gov            & 21   & 9    \\
nih.gov             & 31   & 52   \\
noaa.gov            & 203  & 58   \\
nps.gov             & 13   & 0    \\
nsf.gov             & 2    & 0    \\
osha.gov            & 98   & 62   \\
osmre.gov           & 3    & 0    \\
usda.gov            & 18   & 1    \\
usgs.gov            & 13   & 15   \\
whitehouse.gov      & 3    & 0   
\end{tabular}
\caption{Count of triplets by agency and path depth}
\label{table:final}
\end{table}

%flow chart like anat ben david for source of url

\subsubsection{Provenance of 2008 Mementos}

%https://github.com/lesleyodu/odu-ms-thesis/blob/a1f950b5acf257651343fb74df093c1d43a46d45/pr/2024-03-21/provanalyout.txt
%https://github.com/lesleyodu/odu_phd/blob/main/2024-09-24/provbytld.csv
%save page now?

%all prov
%https://docs.google.com/document/d/1TIw-46M5sUlqU5Uf52eQc8JEzUNtsuNa8C8FpzfQXow/edit?usp=sharing

\begin{figure}[h]
  \centering
  \includegraphics[width=0.7\linewidth]{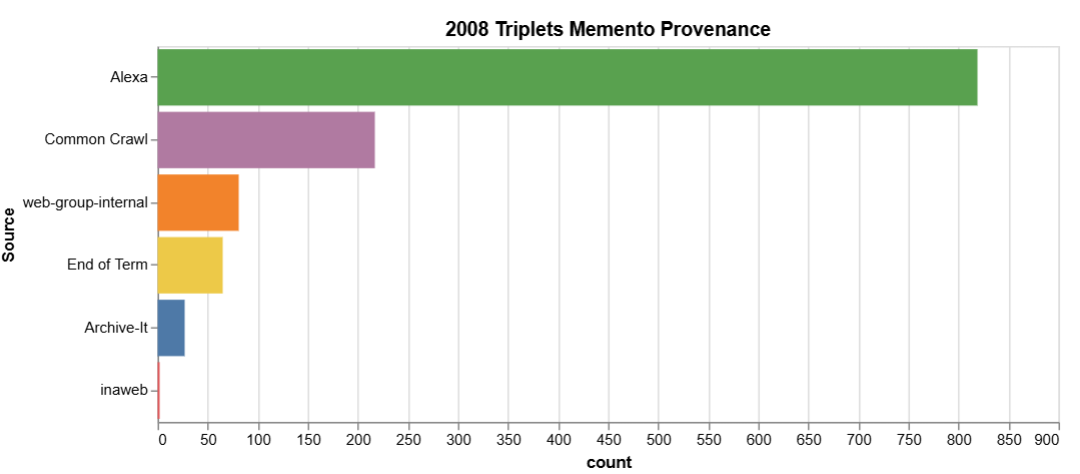}
  \caption{Provenance of 2008 triplet memento by source}
  \label{fig:provchart1}
\end{figure}

Many of the 2016--2020 paired mementos were archived because of EDGI. However, EDGI did not exist in 2008. To analyze the origin of the 2008 mementos, we gathered the provenance information from each capture. Because of rate limiting, one page's provenance can be collected every 15 seconds. Two-thirds (819) of the 2008 mementos were captured by Alexa, as shown in Figure \ref{fig:provchart1} and Table \ref{table:prov}. Of the remaining mementos, 217 were captured by Common Crawl, 81 were captured in internal crawls by the Internet Archive, 65 were captured by the End of Term 2008 crawl, 27 were captured by crawls from Archive-It partners, and 2 were captured by INA, a French national archive.

\begin{table}[]
\begin{tabular}{lllllll}
                & archive it & internal & commoncrawl & inaweb & eot & alexa \\
blm             & 0          & 0        & 3           & 0      & 0   & 4     \\
cdc             & 3          & 0        & 0           & 1      & 0   & 64    \\

doe             & 0          & 0        & 0           & 0      & 0   & 3     \\

doi             & 1          & 2        & 7           & 0      & 0   & 0     \\
dot             & 0          & 1        & 0           & 0      & 0   & 5     \\

energy          & 0          & 0        & 0           & 0      & 0   & 3     \\
epa             & 15         & 10       & 80          & 0      & 45  & 10    \\
federalregister & 0          & 0        & 0           & 0      & 0   & 1     \\
fema            & 1          & 0        & 0           & 0      & 0   & 0     \\

ferc            & 1          & 8        & 0           & 0      & 3   & 169   \\
fws             & 0          & 0        & 14          & 0      & 1   & 75    \\
gao             & 0          & 26       & 0           & 0      & 0   & 1     \\
globalchange    & 0          & 0        & 0           & 0      & 0   & 3     \\
gsa             & 0          & 0        & 1           & 0      & 0   & 1     \\

hhs             & 2          & 0        & 0           & 0      & 0   & 10    \\
justice         & 0          & 19       & 0           & 0      & 0   & 0     \\

nasa            & 0          & 0        & 0           & 0      & 0   & 32    \\
nih             & 3          & 1        & 73          & 0      & 6   & 0     \\

noaa            & 1          & 7        & 28          & 0      & 9   & 218   \\
nps             & 0          & 0        & 0           & 0      & 0   & 13    \\
nsf             & 0          & 0        & 0           & 0      & 0   & 2     \\

osha            & 0          & 0        & 7           & 0      & 0   & 153   \\
osmre           & 0          & 2        & 0           & 0      & 0   & 1    \\

whitehouse      & 0          & 2        & 0           & 0      & 0   & 0     \\
usda            & 0          & 3        & 2           & 0      & 1   & 13    \\
usgs            & 0          & 0        & 2           & 1      & 0   & 38    \\

\end{tabular}
\caption{Count of triplets by agency and 2008 provenance}
\label{table:prov}
\end{table}

\begin{figure}[h]
  \centering
  \includegraphics[width=0.7\linewidth]{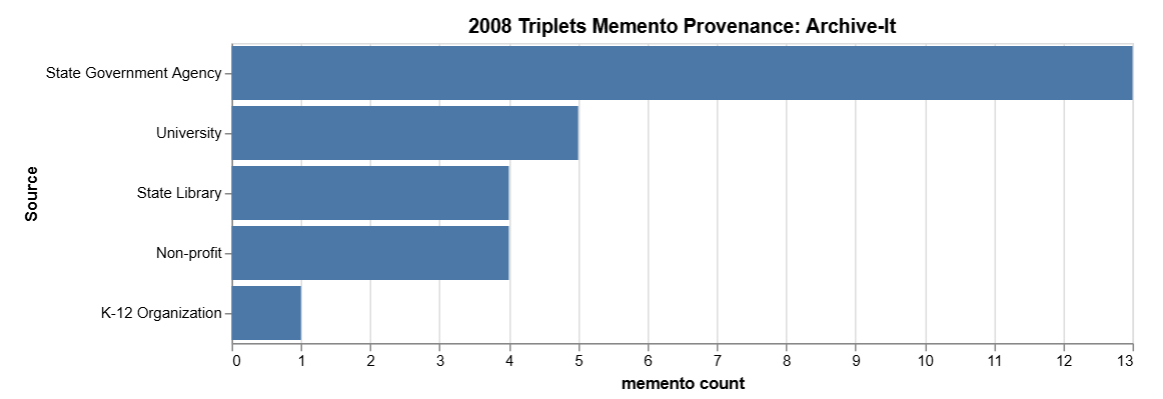}
  \caption{Provenance of 2008 triplet mementos from Archive-It}
  \label{fig:provchart2}
\end{figure}

The Archive-It partners who contributed the specific mementos consist of public and private universities, state government agencies, state libraries, K-12 organizations, and nonprofits, as shown in Figure \ref{fig:provchart2}. All of the mementos from 2008 were saved via crawls, as Save Page Now was not introduced until 2014 \cite{ben2018internet} and Archive Team was only founded in 2009. However, conducting the Archive-It crawls was carried out by nonprofit, commercial, educational, research, and government organizations, and each organization's contribution was important to the overall resulting dataset.

\begin{figure}[h]
  \centering
  \includegraphics[width=0.7\linewidth]{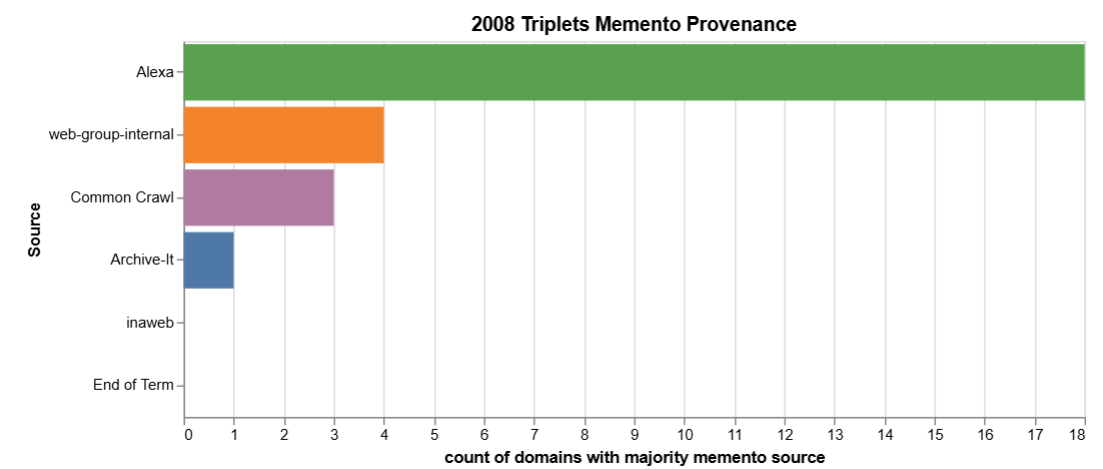}
  \caption{Provenance of 2008 triplet mementos by majority source for each agency}
  \label{fig:provchart3}
\end{figure}

It is also worthwhile to note that aggregating all of the mementos collected in 2008 by non-profits still amounts to only half of the commercial contribution, and the End of Term contribution is only about five percent. In fact, Figure \ref{fig:provchart3} shows that the end of term is not a majority source for any top level domain in the collection. Since the first successful memento from 2008 was saved to the dataset, it is probable that additional captures of the pages were archived by the End of Term crawl later in the year. However, it is still a surprisingly low percent given the topic of the collection. Accordingly, researchers should not rely on only one data source for this early time period, since this dataset could only be assembled via aggregation from 16 organizations.

%maybe a chart with x count high vs y count deep. opportunity for circle size or color to represent another attribute.

\subsubsection{Provenance of Triplets over time}

\begin{figure}[h]
  \centering
  \includegraphics[width=0.7\linewidth]{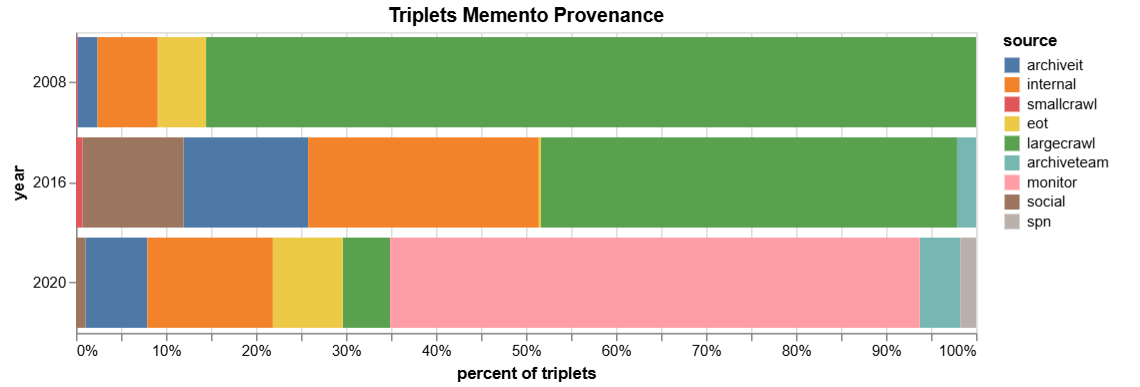}
  \caption{Provenance of triplet mementos over time}
  \label{fig:provchart4}
\end{figure}

Figure \ref{fig:provchart4} shows the distribution of mementos by source over time. Alexa and commoncrawl, both external large crawls, comprise the majority of the sources for the 2008 mementos, while by 2020 this is a very small source. Additional large crawls contributing to the 2020 mementos are perma.cc and the Portuguese Web Archive, as shown in Table \ref{table:provgroup}. Unsurprisingly, monitoring organizations like EDGI make up the majority of memento sources for 2020. While Save Page Now existed by 2016, no mementos had this source during that time period. Archive Team, which was founded after 2008, contributed mementos to the 2016 part of the dataset and even more mementos to the 2020 part of the dataset. Individual organizations archiving through Archive-It continued to contribute meaningful amounts of mementos to the datasets over time.

Each of the nine provenance sources exhibit one of four growth or decay patterns. Always growing, shown stacked towards the right in Figure \ref{fig:provchart4}, includes monitoring organizations, Save Page Now, and Archive Team. These provenance trends for this dataset show the rise in citizen archiving. The rise in archive team contributions and external monitoring groups shows awareness in the archiving community regarding the fragility of webpages in this dataset, and the rise in save page now shows awareness in the general public. Always shrinking includes large imported crawls. The decline in Alexa was perhaps a precursor to the end-of-life of the service in 2022, while Common Crawl's change in priorities towards country level TLDs over generic TLDs around 2016\footnote{https://commoncrawl.github.io/cc-crawl-statistics/plots/tld/groups.html} could explain its decline. The third group follows a grow then shrink pattern, shown towards the left in Figure \ref{fig:provchart4}, and includes internal crawls, Archive-It, small imported crawls, and social media sources. These sources may have a shrink pattern by 2020 only due to the monitoring groups' dominating presence. The last pattern is shrink then grow, for the end of term web crawls. Paired with the internal crawls growing then shrinking, one possible explanation is that the partnership between the Internet Archive and the End of Term crawls shows a change in priority of internal crawls similar to that seen in Common Crawl, with .gov websites being prioritized by End of Term instead of duplicating efforts.

\begin{table}[]
\begin{tabular}{ll}
Source type           & Organizations                                         \\
Large imported crawls & Alexa, Common Crawl, Perma.cc, Portuguese Web Archive \\
Archive-It            & Archive-It                                            \\
Small imported crawls & INA, IMLS, NARA                                       \\
Internal Crawls       & Internet Archive, GDELT (2016)                        \\
End of Term           & End of Term                                           \\
Archive Team          & Archive Team                                          \\
Social                & Wikipedia, Twitter                                    \\
Monitoring            & EDGI, GDELT (2020), Mediacloud                        \\
Save Page Now         & Internet Archive                                     
\end{tabular}
\caption{Grouping of organizations contributing mementos in the triplets dataset}
\label{table:provgroup}
\end{table}

\pagebreak

\subsection{Change Analysis}

\begin{figure}[h]
  \centering
  \includegraphics[width=0.7\linewidth]{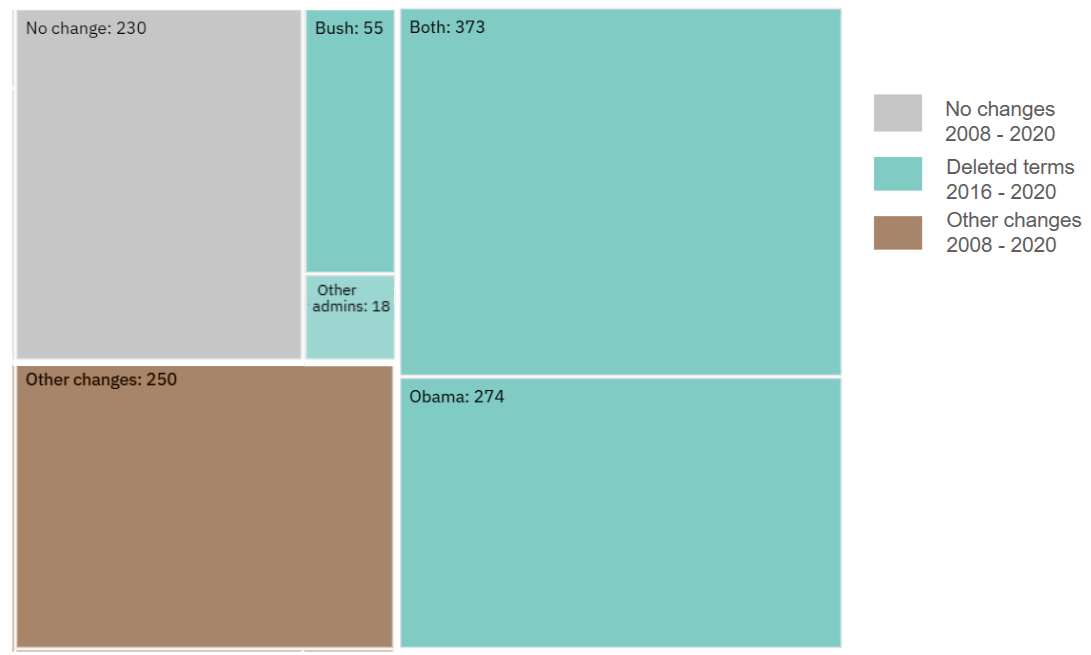}
  \caption{Types of changes by page in the 2008 -- 2020 triplets dataset by presidential administration according to when the deleted terms were added}
  \label{fig:changetypes}
\end{figure}

We analyzed the terms that were fully deleted between July 2016 and July 2020, representing the Trump administration, to determine whether those terms were added before June 2008, representing the Bush administration, or between July 2008 and June 2016, representing the Obama administration. As shown in Figure \ref{fig:changetypes}, out of the 1,220 pages identified for inclusion in the dataset and analysis, we found that 990 pages had changes between 2008 and 2020, and 740 of the pages included terms deleted during the Trump administration. Of those pages, 373 contained deleted terms added during both the Obama and Bush administrations. 274 additional pages had deleted terms only from the Obama administration, representing 37\% of the pages with deletions, while only 55 pages had deleted terms only from the Bush administration. In all, 87\% of the pages with deleted terms had a deleted term that was added during the Obama administration.

%https://www.draxlr.com/tools/tree-map-chart-generator/

%81 percent with changes, this is large for gov, look at top vs deep

%https://github.com/lesleyodu/odu_phd/blob/main/2024-10-01/vout20241001.txt

\subsubsection{Change Trends}

Next, we analyzed deleted terms by agency where the term was deleted during the Trump administration on at least 5 pages on that agency's website, which we refer to as change trends. Certain agencies, such as NIH, NOAA, and OSHA with 105, 88, and 359 change trends respectively, included far more change trend deleted terms than other agencies including EPA, with 20 change trends. This is surpising because of the focus on the deletions on EPA's website by Nost et al. \cite{nost2021} as well as the large number of EPA pages in the triplets dataset. Some examples of climate and regulation change trend deleted terms include ``exposure'' (8 pages on OSHA) and ``healthier'' (5 pages on NIH). 17 agencies included no terms that met the criteria, not even stop words or years. This is most likely explained by the small number of pages on many domains, as shown in Figure \ref{table:final}.

\subsubsection{Emphasis of Climate and Regulation in Deletions}

We analyzed the 56 EDGI tracked terms and phrases to determine which terms added during the Obama administration were commonly deleted. The EDGI terms list includes climate terms, but also includes other terms more focused on regulation. In fact, the most commonly deleted terms belong to the regulation group, such as regulation and safety, rather than the climate group, such as climate and sustainable, as shown in Table \ref{table:termct}.

\begin{table}[]
\begin{tabular}{lll}
Term              & count   & category        \\
regulation           & 16 & regulation \\
safety & 13 & regulation \\
sustainable & 8  & climate \\
emission & 7       & climate   \\
climate              & 5  & climate\\
economic             & 4     & regulation    \\
pollution       & 3      & climate    \\
wildfires & 3 & climate
\end{tabular}
\caption{Most commonly deleted EDGI terms, by category (regulation or climate) showing regulation commonly deleted terms having higher deletion counts than climate commonly deleted terms.}
\label{table:termct}
\end{table}

\subsubsection{Exemplars Case Study: Extension to Biden Administration}

We identified 33 exemplars of the 1,220 pages based on having a deleted EDGI term between 2016 and 2020, as well as a small number of deleted terms added by Bush compared to deleted terms added by Obama. While these pages had persisted from 2008 to 2020, only one-third further persisted until October 2024.

For the pages that decayed between 2020 and 2024, we analyzed whether these pages persisted at a new URL or if they had been deleted, as indicated by a 404 HTTP status code. We found 14 pages with machine readable redirects to go to a new URL. For the 7 pages with a 404 status code, we manually located the new URL for the page manually for 3 pages, another page included the new URL on the 404 page rather than using a machine-readable redirect, and the last 3 pages had a manual redirect by April, 2025, so no pages were truly deleted. Using the redirect categories enumerated by Garg et al. \cite{garg2025websci}, we found both canonical and non-canonical redirects, as shown in Table \ref{table:redirtypes}. The non-canonical redirects featured mostly same-level path changes, such as a subdomain change, or deep to high path changes in line with search engine optimization strategies. We also saw that updating technology, such as moving away from Cold Fusion and ASP.NET, played an important part in the change of URLs. We also found one page, the ``Great Lakes Regional Assessment Report Released'' page on globalchange.gov, redirecting to a top-level landing page rather than a new location for the original content. Upon examining the 2020 memento, we found that it was a soft-404 in that it also redirected to this landing page internally but not in a machine-readable way. This is a ``sink'' page, as other similar deep pages exhibit this same redirection to the same generic upper level page with an associated page deletion, rather than to a new location of the content. We also found 1 erroneous indexed page, in that viewing the mementos manually did not support the change text index calculations. This happens for a variety of reasons, such as when a page includes dynamically loaded content mis-captured during the archival process.

\begin{table}[]
\begin{tabular}{lll}
Redirect type       & Redirect category       & count           \\
Non-WWW to WWW           & Canonical & 4 \\
Old to New Page (3xx) & Non-Canonical & 8 \\
Old to New Page (404) & Non-Canonical & 7 \\
Subdomain Change & Non-Canonical & 2 \\
Sink & Soft 404 & 1
\end{tabular}
\caption{Types of redirects show that most pages have new, non-canonical URLs}
\label{table:redirtypes}
\end{table}

\begin{figure}[h]
  \centering
  \includegraphics[width=0.9\linewidth]{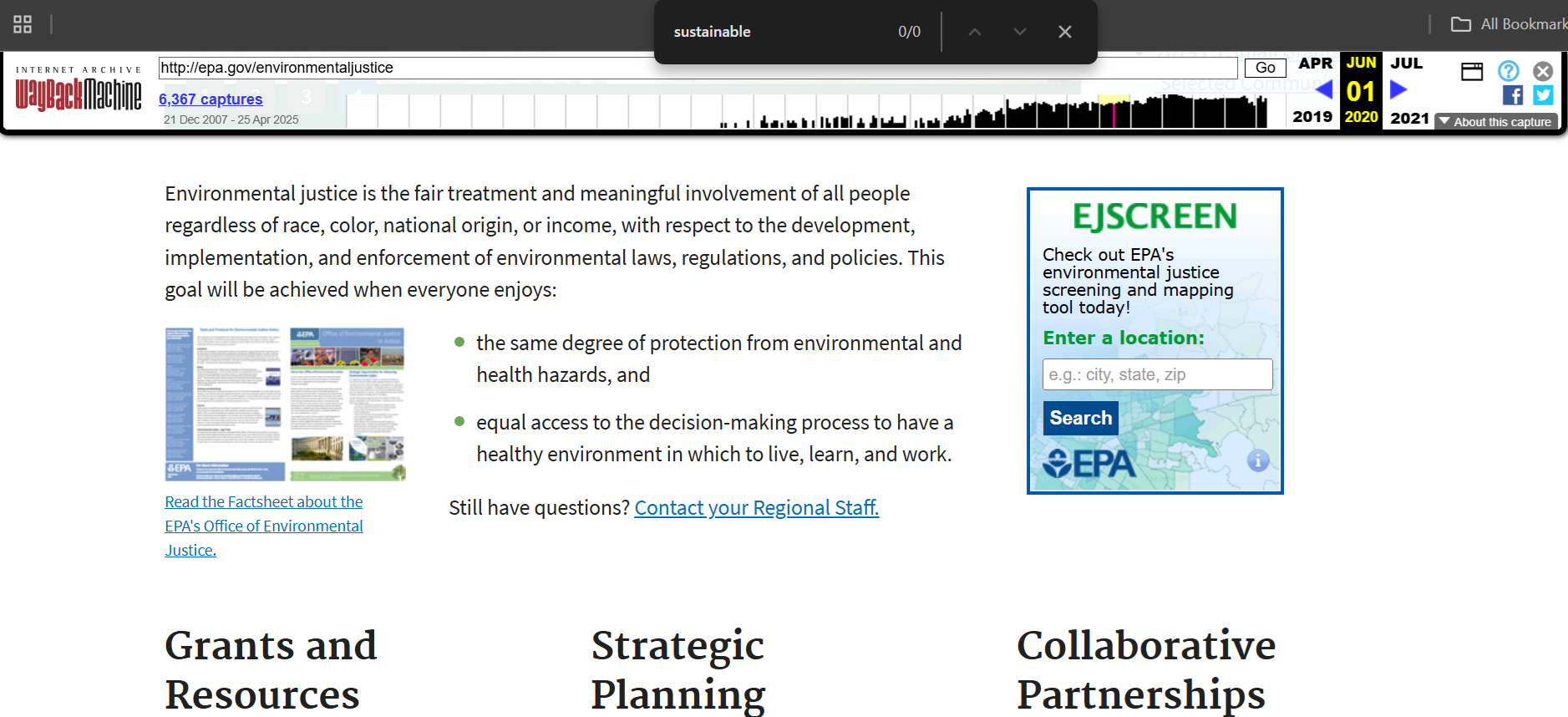}
  \caption{The 2020 version of epa.gov/environmentaljustice had the term ``sustainable'' deleted, which had been added under Obama}
  \label{fig:putback1}
\end{figure}

\begin{figure}[h]
  \centering
  \includegraphics[width=0.9\linewidth]{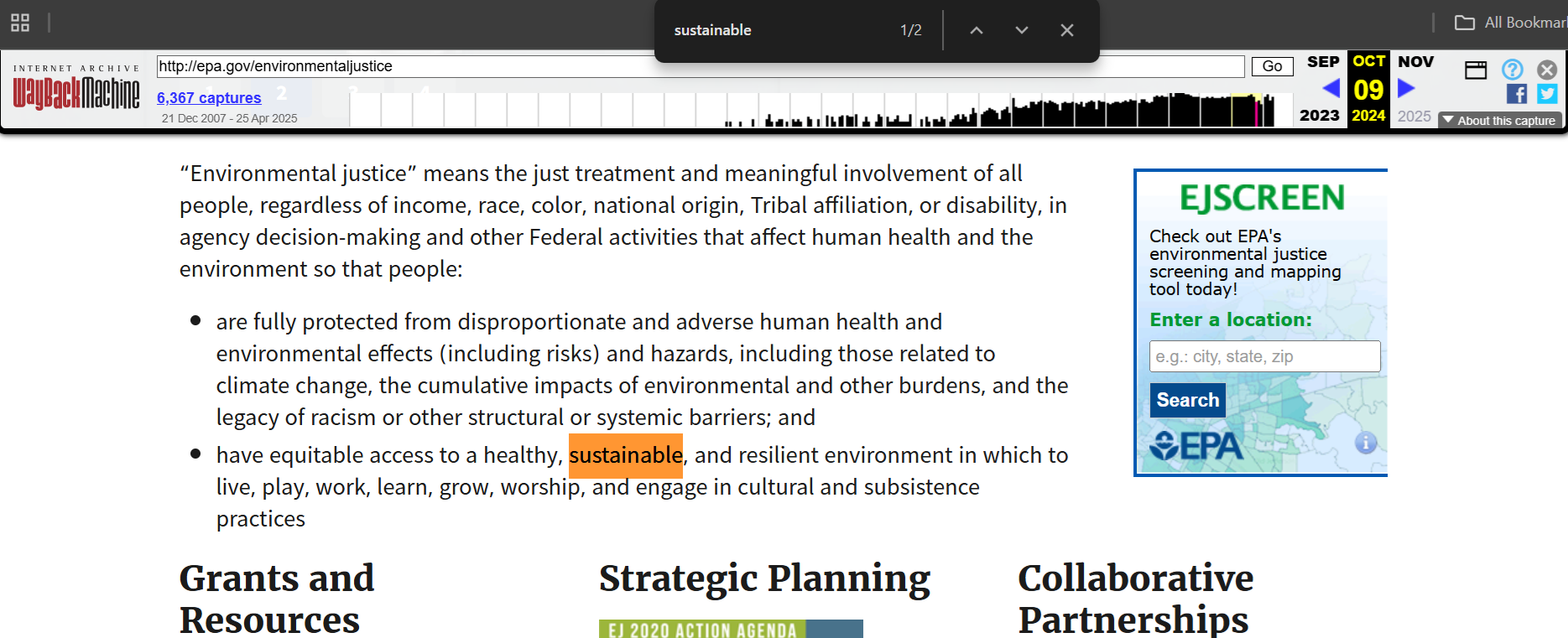}
  \caption{The 2024 version of epa.gov/environmentaljustice restored the deleted term ``sustainable''.}
  \label{fig:putback2}
\end{figure}

\begin{figure}[h]
  \centering
  \includegraphics[width=0.9\linewidth]{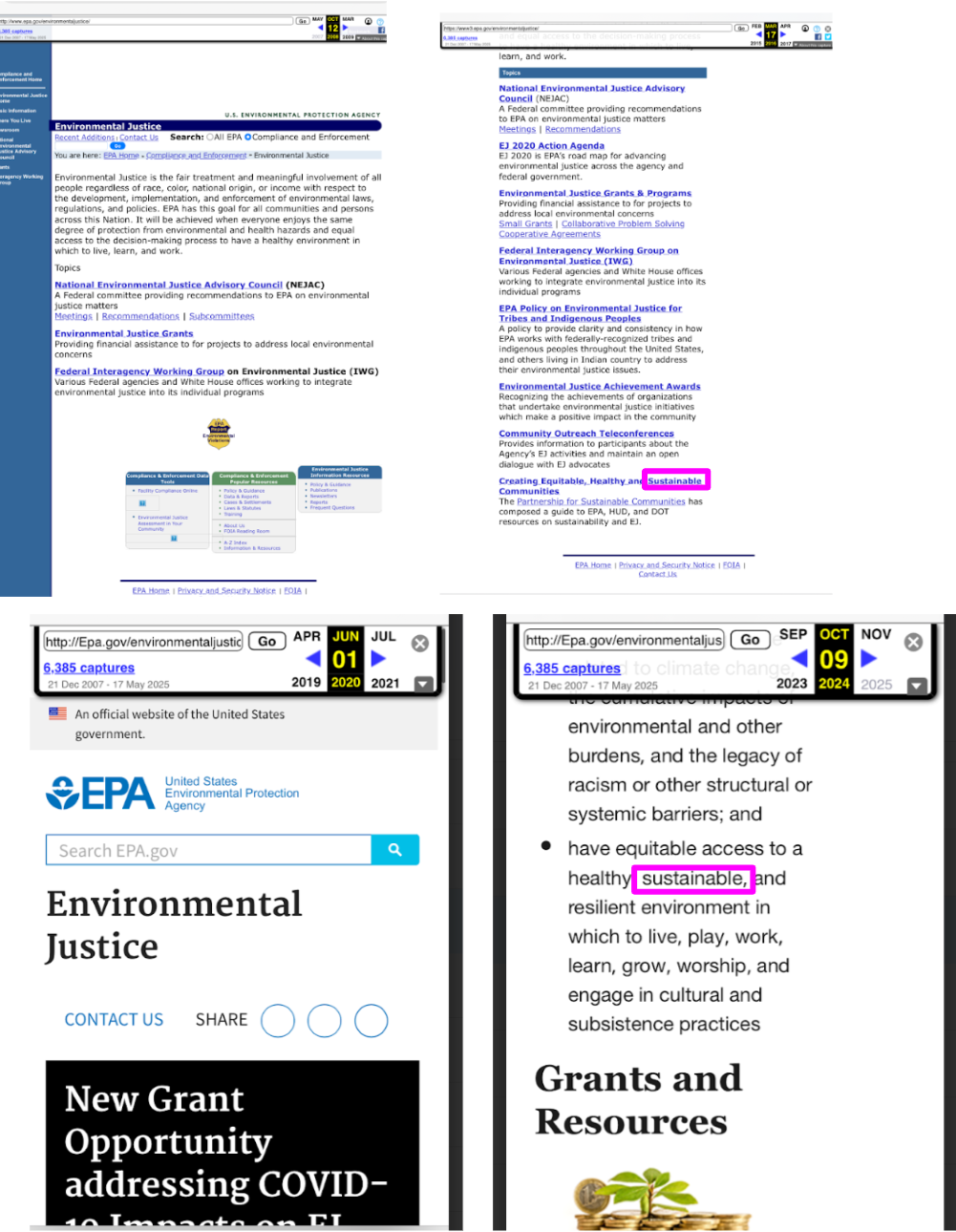}
  \caption{The 2008, 2016, 2020, and 2024 versions of epa.gov/environmentaljustice showing the term ``sustainable'' being added, removed, and restored.}
  \label{fig:compare4}
\end{figure}

For the pages that persisted until 2024, we manually investigated whether any of the deleted terms had been restored by the next presidential administration after Trump. As this administration was the Democratic Biden administration, it is reasonable to examine additions in the same way that we examined deletions for the previous administration due to the opposite political party and thus agenda priorities of the administration. We found 6 pages with restored climate and/or regulation terms and 4 pages with no restored terms. Figure \ref{fig:putback1} shows The EPA's Environmental Justice page missing the term ``sustainable,'' which had been added during the Obama administration. By 2024, the term ``sustainable'' had been restored on the page, as shown in Figure \ref{fig:putback2}. Figure \ref{fig:compare4} shows the four versions of the page spanning 2008 through 2024 side by side along with the term ``sustainable'' indicated clearly on each page version.

%\subsection{Other presidential administrations}
%kritika arxiv preprint had to combine 1997-2000

\clearpage

\section{Conclusions}

In this paper, we outlined a methodology to construct web archive collections for longitudinal analysis after the events have occurred using existing captures aggregated from many sources. We demonstrate the necessity of many types of organizations, from commercial to non-profit along with citizen archiving, in contributing to the final collection. We also outline why automated methods are insufficient to create such a collection. We used our collection to examine changes to webpages over three presidential administrations, and show that the Republican president Trump deleted terms mostly added during the previous Democrat President Obama's administration, highlighting the fluctuation in tenets between the two parties.

\bibliographystyle{ACM-Reference-Format}
\bibliography{myref}

%https://tex.stackexchange.com/questions/20308/creative-commons-logo
\doclicenseThis

\end{document}